\pdfoutput=1
\documentclass[nonacm, sigconf]{acmart}
\renewcommand\footnotetextcopyrightpermission[1]{} 
\usepackage{algorithm, amsmath,amssymb,amsfonts}
\usepackage[noend]{algpseudocode}
\renewcommand{\algorithmiccomment}[1]{\bgroup\hfill\small\##1\egroup}
\usepackage{bm}
\usepackage{float}
\usepackage{graphicx}
\usepackage{multirow}
\usepackage{multicol}
\usepackage{placeins}
\usepackage{subcaption}
\usepackage{textcomp}
\usepackage{hyperref}
\usepackage{xcolor}
\def\BibTeX{{\normalfont B\kern-0.5em{\scshape i\kern-0.25em b}\kern-0.8em\TeX}}

\begin{document}
\title{STAND: A Spatio-Temporal Algorithm for Network Diffusion Simulation}

\author{Fangcao Xu}
\email{xfangcao@psu.edu}
\affiliation{%
  \institution{Department of Geography, Pennsylvania State University}
  \city{State College}
  \state{Pennsylvania}
  \country{USA}
}

\author{Bruce Desmarais}
\email{bdesmarais@psu.edu}
\affiliation{%
  \institution{Depatment of Political Science,
  Pennsylvania State University}
  \city{State College}
  \state{Pennsylvania}
  \country{USA}}

\author{Donna Peuquet}
\email{peuquet@psu.edu}
\affiliation{%
  \institution{Department of Geography, Pennsylvania State University}
  \city{State College}
  \state{Pennsylvania}
  \country{USA}}

\begin{abstract}
Information, ideas, and diseases, or more generally, contagions, spread over time and space through individual transmissions via social networks, as well as through external sources. A detailed picture of any diffusion process can be achieved only when both a detailed network structure and individual diffusion pathways are obtained. Studying such diffusion networks provides valuable insights to understand important actors in carrying and spreading contagions and to help predict occurrences of new infections. Most prior research focuses on modeling diffusion process only in the temporal dimension. The advent of rich social, media and geo-tagged data now allows us to study and model this diffusion process in both temporal and spatial dimensions than previously possible. Nevertheless, how information, ideas or diseases are propagated through the network as an overall \textbf{\textit{spatio}}temporal process is difficult to trace. This propagation is continuous over time and space, where individual transmissions occur at different rates via complex, latent connections. \par

To tackle this challenge, a probabilistic spatiotemporal algorithm for network diffusion simulation (STAND) is developed based on the survival model in this research. Both time and geographic distance are used as explanatory variables to simulate the diffusion process over two different network structures. The aim is to provide a more detailed measure of how different contagions are transmitted through various networks where nodes denote geographic locations at a large scale.
\end{abstract}


\maketitle

\section{Introduction}
With the advent of rich social and other web-based media containing both temporal and locational information, tracing the diffusion of ideas, political opinions and even evidence of processes such as the spread of disease at a large scale has become a focus of research in recent years. Diffusion is the process by which contagions spread over space and time via complex network structures. Contagions start at specific nodes and spread from node to node over the edges of the network. These traces are called cascades. \par

Previous network research \cite{liben2008tracing} indicates that observing when individual nodes in the network get infected by various contagions is easy, but determining the transmission pathways is difficult. In other words, the times at which nodes get infected are noted in the observational data but the sequence and parent node through which each node gets infected is usually not. Many algorithms have been developed to simulate cascades for different contagions to drive a better understanding of diffusion process. However, most current algorithms only exploit time as an explanatory variable and don't take the geographic space  into consideration. \par

Nevertheless, all diffusion processes that involve physical agents, locations or interactions, are embedded in geographic space. For examples, news events usually occur at specific locations. People receiving and exchanging information or ideas via social media also have a physical location. or physically being proximal to each other. When we trace the diffusion phenomena at a large scale over a long time, it has also been demonstrated the spread and adoption of many contagions are different from region to region with significant local characteristics \cite{liben2005geographic, leskovec2014geospatial}. Instead of geographic distance becoming increasingly irrelevant, it's more accurate to say technology has made border less relevant.\par

The motivation of this research is to provide a well-defined and mathematically solid approach for solving diffusion simulation and modeling problems taking both spatial and temporal information into consideration. To achieve this, we have developed a probabilistic algorithm called STAND,using a survival approach, to simulate spatiotemporal diffusion cascades. This algorithm is applicable to various types of network structures. It is intended that our research can lead to new insights of how different contagions, including topics, opinions,sentiments, or events mined from world wide web are propagated over space and time. 

\section{Related Work}
Simulating the complete topology of spatiotemporal networks and multiple contagions spread over them is challenging for two reasons: First, in many cases we can only observe the timing information of when nodes get infected \cite{rodriguez2011uncovering}. Second, even though large amounts of digital heterogeneous data are now available via the World Wide Web, locational information is extremely sparse, unstructured and often ambiguous \cite{wang2016spatial,almquist2018large,kabir2019analysis}. Developing a flexible model for deriving the network structure and cascade behaviors is key to uncovering the mechanism that governs spatiotemporal diffusion processes and their dynamics. \par

Several differential equation models (DE) and agent-based models (AB) have been proposed to simulate the network structure and spatiotemporal diffusion cascades. DE models \cite{mahajan2000new,rahmandad2008heterogeneity} usually aggregate agents into several states (e.g., infected or uninfected). The transitions among different states are modeled by differential equations. In contrast, agent-based models \cite{barrett2008episimdemics, Kiesling2012} have considered the heterogeneity of agents. They simulate the diffusion in realistic networks by defining how the infection may occur through individual-based interactions (e.g., infection can only occur when agents are at the same location). \par

While, most existing simulation models of network diffusion are based on assumptions of agent homogeneity or how agents interact with each other for spreading contagions, none have integrated space and time together to account for diffusion probability individually for each node-pair over spatiotemporal networks at large-scale. We also intend to fill this gap. We develop an exponential diffusion probabilistic algorithm that integrates geographic distance, node infection time and transmission speed together based upon NETINF \cite{rodriguez2011uncovering}. The reasons to choose an exponential network diffusion model as a starting point for STAND are: 1) it's a continuous-time model without any assumptions about the individual interactions, 2) it is amenable to inference/estimation for different datasets, and 3) it has a flexible structure in terms of adding features like geo-distance. Other time distributions (e.g., log-normal, Gamma, etc.) can also be considered in the future research as well.

\section{The STAND Algorithm}
Table I defines the parameters used in many diffusion network models, and we build upon this notation. When a specific contagion spreads over the network, it will create a cascade by infecting nodes in a temporal sequence. The cascade of a contagion $c$ is denoted by a directed tree $T$, consisting of a set of infected nodes with observed infection time: $(i,t_i)_c$ where $i \in V$.
\begin{table}
\begin{center}
\begin{footnotesize}
\caption{Parameters in Diffusion Networks}
\begin{tabular}{ l|l }
\hline
\bf Parameters & \bf Def\\
\hline
$G(V,E)$ & Directed Graph with node set V and edge set E\\
$c$ & Contagion that spreads over $G$\\
$C$ & Set of contagions $c$\\
$T$ & Cascade propagation tree\\
$T_c(G)$ & Set of all possible cascade trees of the contagion $c$\\
$t_i$ & Time when node $i$ get infected by a contagion\\
$\Delta_{i,j}$ & Time difference between the node infection time $t_j-t_i$\\
$\alpha$ & Diffusion speed scaling parameter\\
$\beta$ & Probability that contagion spreads over the edge of $G$\\
\hline
\end{tabular}
\end{footnotesize}
\end{center}
\end{table}

\subsection{Temporal Probabilistic Survival Likelihood}
The likelihood of a contagion $c$ spreading over an edge $(i,j)$ in NETINF is calculated by a probabilistic exponential function and assumed to depend only on the time difference $\Delta_{i,j}$ in Equation (1). \par

\begin{equation}\label{eq1}
    P_c(i,j) \approx P_c(\Delta_{i,j}) \propto \frac{\beta}{e^{\alpha \Delta_{i,j}}}
\end{equation}

Only the first time when a node gets infected will be counted for each contagion $c$ and the contagion must diffuse forward in time. If there is a node $j$ that never got infected by any contagion, then $t_j = \infty$ and $\Delta_{i,j} \approx \infty$. Thus each infected node will only have one parent node who diffuses the contagion $c$ and the $t_j > t_i$. \par

However, there may exist more than one possible diffusion cascade tree for the same set of infected nodes. These trees have the same temporal sequence for the occurrences of infections among nodes but their diffusion paths are different. Equation (1) which only exploits the infection time difference to explain the diffusion probability, cannot solve which diffusion path has a higher probability. \par

For example, Figure 1 shows three different trees with the same observed spatiotemporal diffusion sequence where the nodes are geographic locations, $(t_{Washington}, t_{Philadelphia}, t_{Pittsburgh}, t_{Chicago})_c$. The infection time $t_{Washington} < t_{Philadelphia} < t_{Pittsburgh}< t_{Chicago}$. The diffusion path from Washington to Chicago can be $(Washington \rightarrow Philadelphia \rightarrow Pittsburgh \rightarrow Chicago)$ as shown by the red line in the left plot or $(Washington \rightarrow Pittsburgh \rightarrow Chicago)$ in the middle plot.

\begin{figure}
\includegraphics[width=\linewidth]{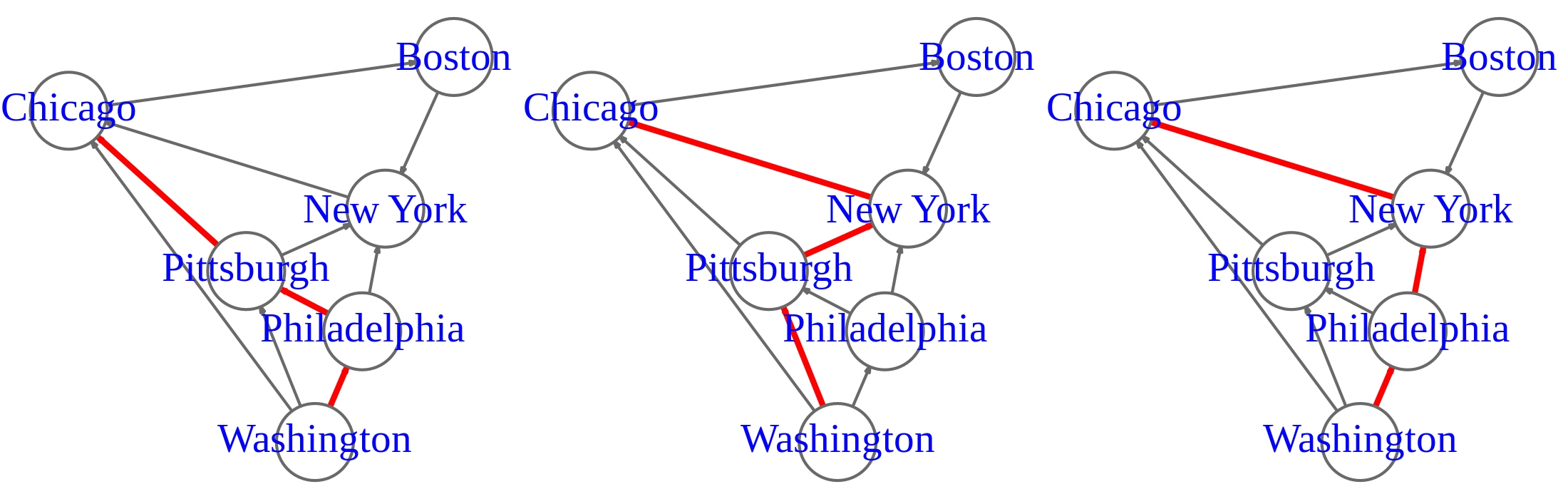}\par
\caption{Different cascade trees} 
\label{fig1}
\end{figure}

The diffusion probability from Washington to Chicago of these three paths defined by Equation \ref{eq1} are same, demonstrated as below:
\begin{equation}
\begin{split}
P_c(i, k, p, j) & =\frac{\beta}{e^{\alpha\Delta_{i,k}}}*\frac{\beta}{e^{\alpha\Delta_{k, p}}}*\frac{\beta}{e^{\alpha\Delta_{p,j}}}  = \frac{\beta^3}{e^{\alpha(\Delta_{i,k}+\Delta_{k,p}+\Delta_{p,j})}} \\
& = \frac{\beta^3}{e^{\alpha\Delta_{i,j}}} = \frac{\beta^3}{e^{\alpha(\Delta_{i,p}+\Delta_{p, q}+\Delta_{q,j})}} = P_c(i, p, q, j) \\
& = \frac{\beta^3}{e^{\alpha(\Delta_{i,k}+\Delta_{k, q}+\Delta_{q,j})}} = P_c(i, k, q, j)
\end{split}
\end{equation}
where $i$ is Washington, $k$ is Philadelphia, $p$ is Pittsburgh, $q$ is New York and $j$ is Chicago.\par

It's thus needed to extend the current diffusion probabilistic algorithms to include space as well as time to better model the dynamic diffusion over the spatiotemporal network.

\subsection{Spatiotemporal Probabilistic Survival Likelihood}
We assume the spatiotemporal diffusion probability $P_c(i,j)$ that a cascade $c$ will spread from a node $i$ to a node $j$ decreases with both the spatial distance $d_{ij}$ and time difference $\Delta_{i,j}$ in a certain way. Geographic distance is integrated into Equation (1) as a starting point. Considering that the geographic distance may have direct influence on the diffusion speed and infection time interval, this influence is modeled by the survival analysis model.

Supposing $T$ is the observed infection time of a node, then the probability that this node would not yet be infected at any time $t$ is denoted by the survival function, $S(t) = P(T > t)$.\par

The probability that a given node would get infected at any time $t$, is denoted by a cumulative probability density function (CDF), $F(t) = P(T \leq t) = 1-S(t)$. The probability that a node will get infected within a time interval $(t, t+dt)$ is $P(t \leq T \leq t + dt) = f(t)dt$ where $f(t)$ is the infection rate over time, given by the probability density function (PDF):
\[f(t) = \frac{d}{dt}F(t)=\frac{d}{dt}(1-S(t))=-\frac{d}{dt}S(t) \tag{3}\]

The instantaneous infection rate $h(t)$ at the given time $t$, is called the hazard rate, that reflects the likelihood that the uninfected node will get infected within a very short time interval $(t, t + dt)$, given this node hasn't been infected before: 
\[h(t) = \lim\limits_{dt \to 0}\frac{Pr(t \leq T < t+ dt)}{dt \times S(t)} = \frac{f(t)}{S(t)}\tag{4}\]

The survival likelihood $S(t)$ and the cumulative infection probability $F(t)$ have the following relationships with the hazard rate $h(t)$: 
\begin{align}
    S(t) = e^{-\int_{0}^{t}h(t)dt} && F(t)=1- e^{-\int_{0}^{t}h(t)dt} \tag{5}
\end{align}
where the $\int_{0}^{t}h(t)dt$ is the cumulative hazard that represents the total risks of a node being infected up to the time point $t$.\par

In order to measure pairwise infection likelihood in Equation (1), the probability density $f(t)$ and the hazard rate $h(t)$ are needed to calculate the probability that a node $j$ gets infected by the node $i$ within the time interval $(t_i,t_i+dt)$. A higher value of density function $f(t)$ means the node $j$ is more likely to be infected by the node $i$ within a short time interval $dt$ which results in a higher hazard rate.\par

In this step, we use the a proportional hazard function suggested by \cite{taylor2017spatsurv} to exploit the geographic properties in the form below:
\[h(t;Y;\lambda;\theta) = e^{(\lambda X + Y)} h_0(t;\theta) \tag{6}\]
where the $h_0(t;\theta)$ is the baseline hazard function, and $t$ is the observed infection time. $X$ is a vector of explanatory variables associated with node properties and $Y$ is a vector of explanatory variables associated with the spatial dependencies. $\lambda$ and $\theta$ are parameter coefficients for the explanatory variables $X$ and the baseline hazard function. \par

To simplify, the baseline hazard rate $h_0(t;\theta)$ in Equation (6) is assumed to be constant and independent with time for any node-pairs. Thus the $h_0(t;\theta)$ and its associated cumulative hazard rate $H_0(t;\theta)$ are:
\begin{align}
   h_0(t;\theta) = \theta && H_0(t;\theta)= \int_{0}^{t}h_0(t;\theta)dt = \theta t \tag{7}
\end{align}

Furthermore, only the spatial distance is implemented into the diffusion probability as a starting point. Thus the $\lambda X$ is a constant modeled by the parameter $\lambda$ and the explanatory variables $Y$, accounting for spatial dependencies in Equation (6) is modeled by the distance between any node-pair $i$ and $j$ with a decay parameter. To unify the parameter coefficients, the $\lambda$ for node properties is reset as $\lambda_0$ and the decay parameter for spatial dependencies (i.e., distance here) is set as $\lambda_1$. Equation (6) can be reformatted as:
\[h(t;Y;\lambda;\theta) = \theta e^{(\lambda_0 + \lambda_1d_{i,j})} \tag{8}\]
where the $d_{i,j}$ is the geographic distance between the node $i$ and $j$. \par

The cumulative probability $F(t)$, the probability density $f(t)$, and the infection likelihood defined in Equation (1) with the hazard rate $\theta e^{(\lambda_0 + \lambda_1d_{i,j})}$ over the infection time interval $\Delta_{i,j}$ can be rewritten as:
\begin{equation}\tag{9}
\begin{split}
&F_{i,j}(\Delta_{i,j})= 1- e^{-\int_{0}^{t}h(t)dt} = 1- e^{-\theta e^{(\lambda_1 + \lambda_1d_{i,j})}\Delta_{i,j}} \\
&f_{i,j}(\Delta_{i,j}) = \frac{\theta e^{(\lambda_0 + \lambda_1d_{i,j})}}{e^{\theta e^(\lambda_0 + \lambda_1d_{i,j}) \Delta_{i,j}}} \\
&P_c(i,j) \propto \beta f_{i,j}(\Delta_{i,j})= \frac{\beta\theta e^{(\lambda_0 + \lambda_1d_{i,j})}}{e^{\theta e^(\lambda_0 + \lambda_1d_{i,j}) \Delta_{i,j}}} 
\end{split}
\end{equation} 

The network inference problem can also be solved by developing a greedy algorithm to search for all edges that can maximize the probability of a network structure over which all contagions $C$ can spread.

\section{Experimental Simulation}
In order to simulate real-world diffusion processes, and evaluate our algorithm for recovering the network structure on the synthetic cascades, experiments are described in the following sections.

\subsection{Simulation of Ground-truth Network}
One hundred geographic cities/counties with the largest population in the United States, excluding Hawaii and Alaska, are taken as spatial network nodes $V$. The original data is from the U.S. Geological Survey collected in 2017. Some cities/counties having high population and very close geographic locations were merged into larger metro areas (e.g., New York City or Greater Los Angeles). Then we manually deleted a few small cities and add other cities in Montana, Wyoming, North and South Dakota to make the nodes more evenly distributed over the United States.\par

Considering that many real-world networks share some similar properties, such as relatively small average path length, high clustering coefficient, or high reciprocity (i.e., the percentage of mutually connected node-pairs), two directed network structures are considered to generate the ground-truth network $G$. They are the random network structure \cite{erdos1960evolution} and the small-world network structure \cite{watts1998collective}. Characterizing how contagions spread over these two extreme network structures can help us better understand more complex network diffusion phenomena. However, our proposed STAND algorithm can be applied to any kind of network structure. The pseudocode of STAND algorithm is given as below:

\begin{algorithm} 
\caption{STAND Algorithm}
\begin{algorithmic}
\Require coordinates of $n$ nodes $(lng, lat)$, number of edges $m$, network $G(V,E)$, parameter $\beta, \theta, \lambda_0, \lambda_1$, cutoff diffusion time $t_{max}$, number of nearest neighbours $k$.
\State \textbf{Step 1: Simulate Random \& Small-world Networks}
\State $V \gets (lng, lat)$  
\State $G_{random} \gets E \gets COMBINATION(V,2,m)$  
\ForAll{node $v \in V$}
\State $G_{sw} \gets E \gets (v, K\_NEAREST(v,k, Euclidean))$  
\State $RECONNECT(E \in G_{sw}(V,E), 0.05)$
\EndFor
\State \textbf{return} $G_{random}, G_{sw}$
\State \textbf{Step 2: Simulate Cascades}
\ForAll{edges $(i,j) \in G_{random}|G_{sw}$}
\State $d_{i,j} \gets EuclideanDist ((lng_i, lat_i),(lng_j,lat_j))$ 
\State $h_{random}|h_{sw} \gets h(i,j) = \theta e^{(\lambda_0 + \lambda_1d_{i,j})}$ \hfill \small{Equation (8)}
\EndFor
\State \textbf{return} $h_{random}, h_{sw}$
\State $C_{random} = SIMULATE(G_{random}, h_{random})$ \hfill \small{Equation (9)}
\State $C_{sw} = SIMULATE(G_{sw}, h_{sw})$
\State \textbf{return} simulated cascades $C_{random}, C_{sw}$
\end{algorithmic}
\end{algorithm}

\subsubsection{Simulation of Random Network}
The random network assumes the node degree in the network has a normal distribution and edges are randomly picked with a constant probability with a same node-edge density. This has been widely used for network simulations with the absence of topological information about the network structure. \par

Building this type of network begins with $n$ spatial isolated nodes and selecting out of those any two to randomly place an edge between them until all edges of the required number have been added to the network without repetition. An example is shown in Figure 2 (Left). 

\begin{figure*}
\begin{multicols}{2}
\includegraphics[width=\linewidth]{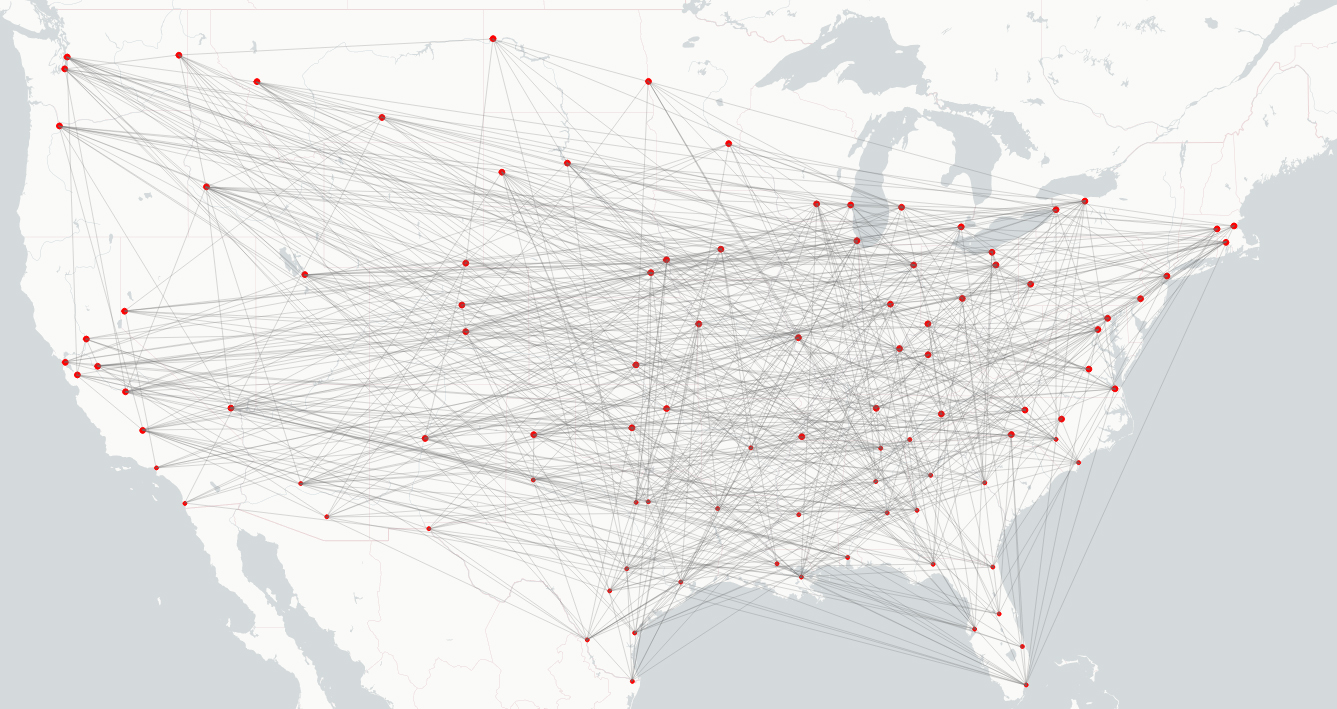}\\
\includegraphics[width=\linewidth]{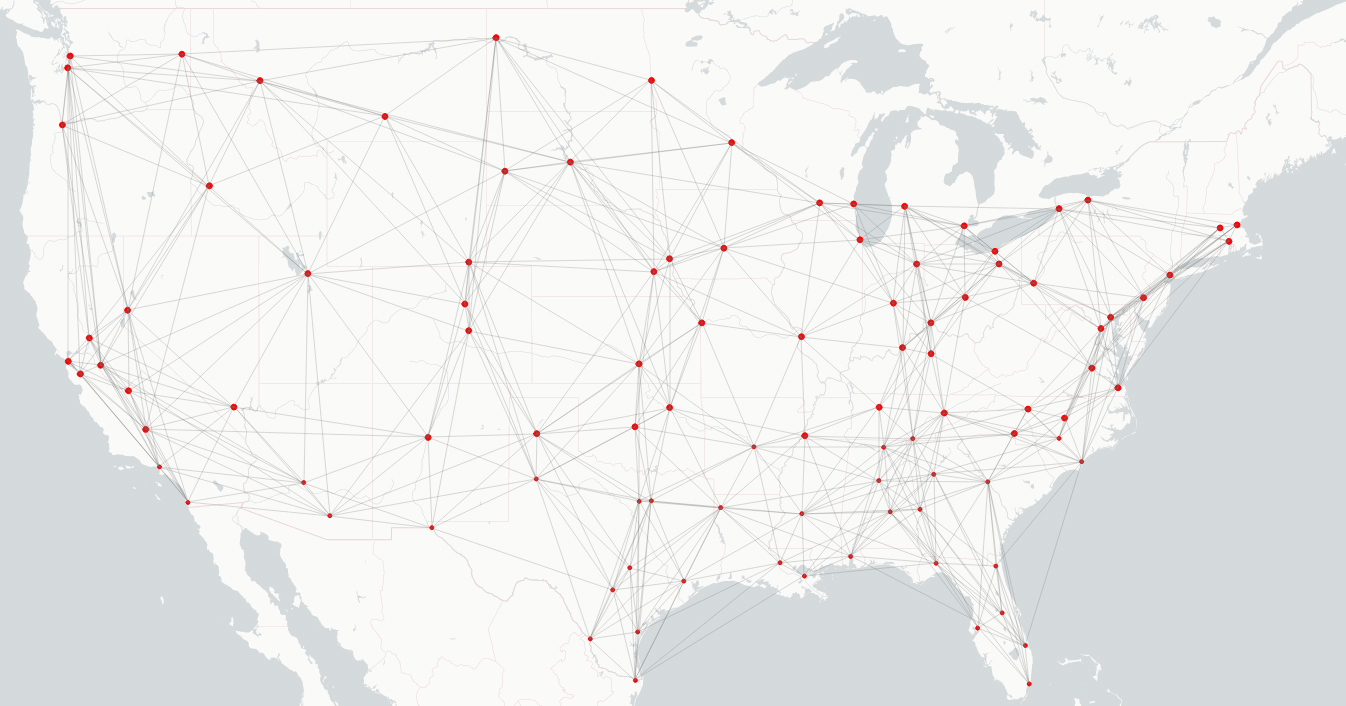}
\end{multicols}
\caption{Random network (Left) and Small-world network (Right) with 100 nodes and 600 edges.}
\end{figure*}

\subsubsection{Simulation of the Small-world Network}
The small-world network as shown in Figure 2 (Right) assumes the node degree follows the power-law distribution, in which only a small proportion of nodes has a large number of links while the majority has only few or no links. Thus it has a high clustering coefficient and a relative short average path length. The small-network model is widely used in many application contexts, such as the outbreak of disease or social activities. \par

Building this type of network starts with connecting each spatial node to its $k$ nearest neighbors. Next, a small proportion of edges are randomly rewired by removing the original edge and reconnecting the starting node of the original edge to another node with a probability $p$ (i.e., 0.05). \par

Here, we have built 600 edges among these geographic locations for each network structure. In comparing left and right figures in Figure 2, it is obvious that the average edge distance of the random network is much longer than that of the small-world network, which has made the small-world network seem sparser over space than the random network even though they have the same number of edges. The small-world network also shows a significant clustering pattern of cliques in contrast to the random network.

\subsection{Simulation of Cascades}
A set of cascades $C$ that spreads over the network $G$ are generated by the probabilistic function defined in Equation (9). At this step, each node is selected as the single starting infected node with assigned an infection time 0, and then it starts to spread the contagion to every remaining node. The infection time of remaining nodes is calculated based on Equation (9). If there is no edge between any two nodes, the hazard rate $\alpha$ is forced to be set as 0.  \par

Different parameter values are selected for generating diffusion cascades. Recalling that the hazard rate $\alpha = \theta e^{(\lambda_0 + \lambda_1d_{i,j})}$ in Equation (8) controls the infection rate that a contagion spreads over edges, there are three parameters to express this: $\theta$ (i.e., a baseline hazard rate), $\lambda_0$ (i.e., influence of node properties), and $\lambda_1$ (i.e., parameter coefficient of influence of distance). \par

Several parameters are set as fixed values in this step because they are constant scaling factors irrelevant to account for the influence of explanatory variables (i.e., the distance). They are: 1) the prior probability of an edge to successfully spread the contagion, $\beta = 0.5$. A higher value of $\beta$ means most of edges are more likely to spread the contagion, which results in large infections/cascade size within the network; 2) the baseline hazard rate, $\theta = 0.5$; 3) the influence of node properties on the diffusion hazard rate, $\lambda_0 = 1$; and 4) the penalty parameter that accounts for the small probability that the contagion may spread via non-existing edges, $\epsilon = 10^{-4}$.\par

The parameter coefficient of the influence of the distance $\lambda_1$ and the cutoff diffusion time $t_{max}$ used to decide to what extent the diffusion should fail when the diffusion time along an edge is too long, are chosen from a list of values in Table II. \par

\begin{table}
\begin{center}
\begin{footnotesize}
\caption{Parameters of $\lambda_1$ and $t_{max}$}
\begin{tabular}{ c|l|l|l|l }
\hline
\multirow{3}{*}{$\lambda_1$} & $-5*10^{-5}$ & $-3*10^{-5}$ & $-2*10^{-5}$ & $-10^{-5}$\\\cline{2-5}
& $-9*10^{-6}$ & $-8*10^{-6}$ & $-7*10^{-6}$ & $-6*10^{-6}$  \\\cline{2-5}
& $-5*10^{-6}$ & $-4*10^{-6}$ & $-3*10^{-6}$ & $-2*10^{-6}$ \\
\hline 
\multicolumn{5}{c}{}\\
\hline
\multirow{2}{*}{$t_{max}$} & 500 & 1000 &  2000 & 4000 \\\cline{2-5} & 8000  & $\infty$ &&\\
\hline
\end{tabular}
\end{footnotesize}
\end{center}
\end{table} 

These values of $\lambda$ and $t_{max}$ increase monotonically and can represent cascades ranging from small (i.e., contagions can almost not spread) to large (i.e., a large infection over all nodes). The reason to set $\lambda_1$ as negative is that in general, the hazard rate $\alpha$ should be small when the spatial distance is very large. In other words, the diffusion time that a contagion spreads via a network edge is longer when two nodes are far away from each other if all other parameters are fixed. The $\alpha$ has a positive relationship with $\lambda_1$. When $\lambda_1$ increases from the $-5*10^{-5}$ to $-2*10^{-6}$, the diffusion/infection speed via network edges will be larger. \par

The cutoff diffusion time $t_{max}$ is used to set those nodes that are infected beyond the threshold as $\infty$.  For example, if 1000 (time units) is selected as the cutoff value, then nodes with simulated infection time more than 1000 should be set as "won't get infected". It can model phenomena that the outbreaks of some contagions happen within a short period then fades very quickly due to geographic constraints while others can easily spread at a large scale and last for a long time. \par

The analysis of the simulated cascades of the static random network and the static small-world network is given in the following section using the parameters selected above.

\section{Analysis of Results}
\subsection{Distribution of the Distance}
The histograms and Quantile-Quantile (QQ) plots in Figure 3 are used to examine the statistical distribution of the edge distance. The range of the statistical distance (km) between two nodes in the random networks is much larger than that in the small-world network. Both of them have a bell-shaped curve as shown in the top of Figure 3. However, the density curves in black suggest that the distribution of the edge distance in the random network is smoother and more widely distributed while the edge distance the small-world is dominated by more lower values. The reason accounting for this in the random network may be due to the limited number of geographic points, thus few observations can easily skew the distribution. For the small-world network, the reason is the existence of a small portion of edges that connect different cliques.\par

\begin{figure}
\includegraphics[width=\linewidth]{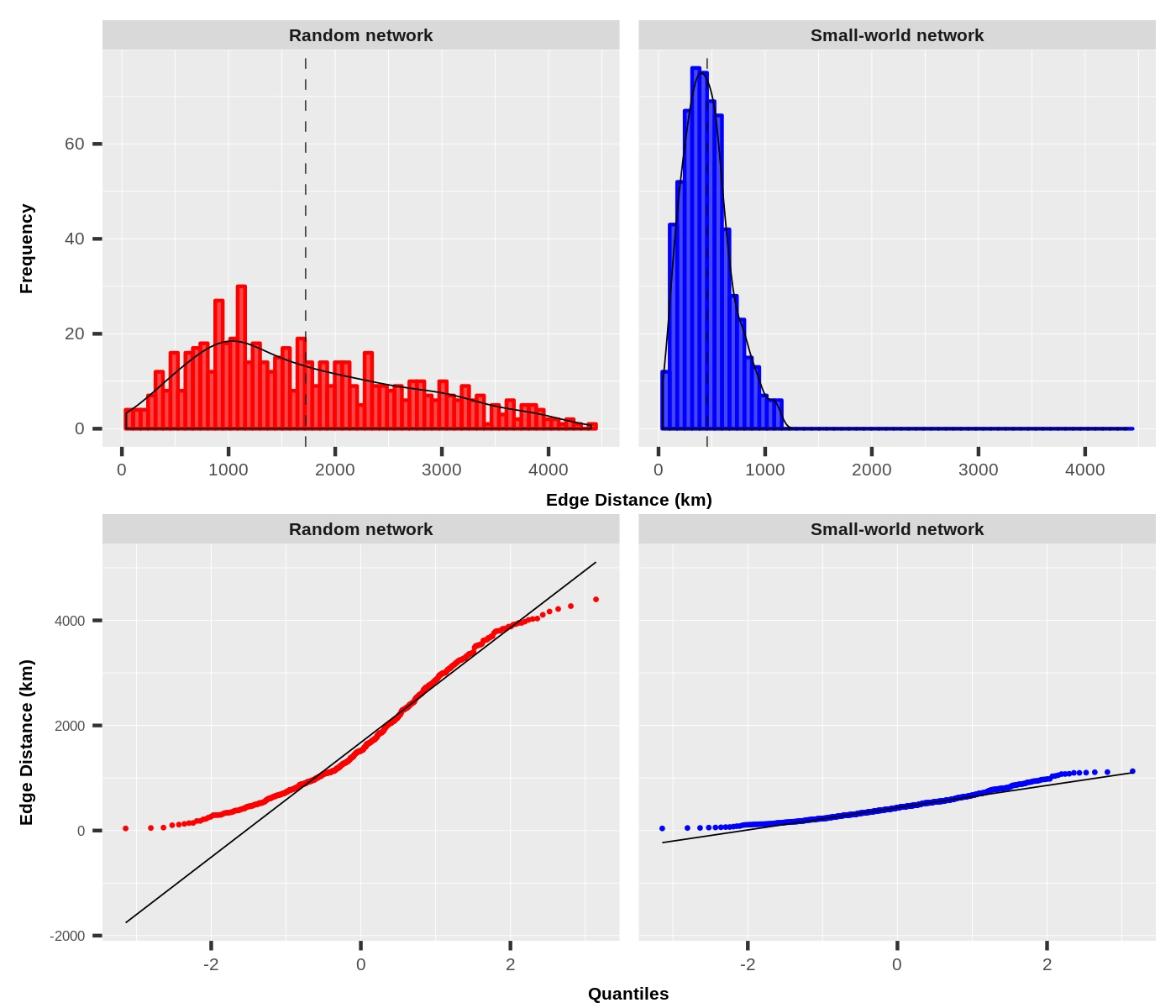}
\caption[Histogram and QQ-plot of the edge distance]{Histogram and QQ-plot of the distance between any two nodes in the Random and Small-world networks}
\vspace{-4mm}
\end{figure} 

\subsection{Infected Nodes in Simulated Cascades}
For each combination of $\lambda_1$ and $t_{max}$ shown in Table II, 100 cascades are generated by taking every node as a starting infected node and the information including the infected node name, the infection time, the parent node from whom the contagion spreads, the diffusion pathway and the cascade id are reported. For example, the information in Table III suggests a diffusion cascade that starts from the node Chicago spreads via two pathways $(Chicago \rightarrow Milwaukee)$ and $(Chicago \rightarrow Fort\,Wayne \rightarrow Grand\,Rapids)$. \par

\begin{table}
\begin{center}
\begin{footnotesize}
\caption{Information about Simulated Cascades}
\begin{tabular}{ l|l|l|l|l }
\hline
node\_name & time & parent\_node & v\_path & cascade\_id \\ 
\hline
3: Chicago & 0.00000 & 3 &  3 & 3   \\
31: Milwaukee & 1.71148 & 3 & (3, 31) & 3 \\
56: Fort Wayne & 3.61669 & 3 & (3, 56) & 3 \\
82: Grand Rapids & 4.03627 & 56 & (3, 56, 82) & 3\\
\hline
\end{tabular}
\end{footnotesize}
\end{center}
\end{table} \par

The box-plots in Figure 4 give a clearer picture of the distribution of the number of infected nodes under different sets of $\lambda_1$ and $t_{max}$. The left series of plots in Figure 5 shows how the total number of infected nodes changes for different values of $\lambda_1$, fixing the cutoff diffusion time while the right plot in Figure 5 shows how the total number of infected nodes changes for different cutoff diffusion time $t_{max}$, fixing $\lambda_1$. \par

\begin{figure}
\includegraphics[width=\linewidth]{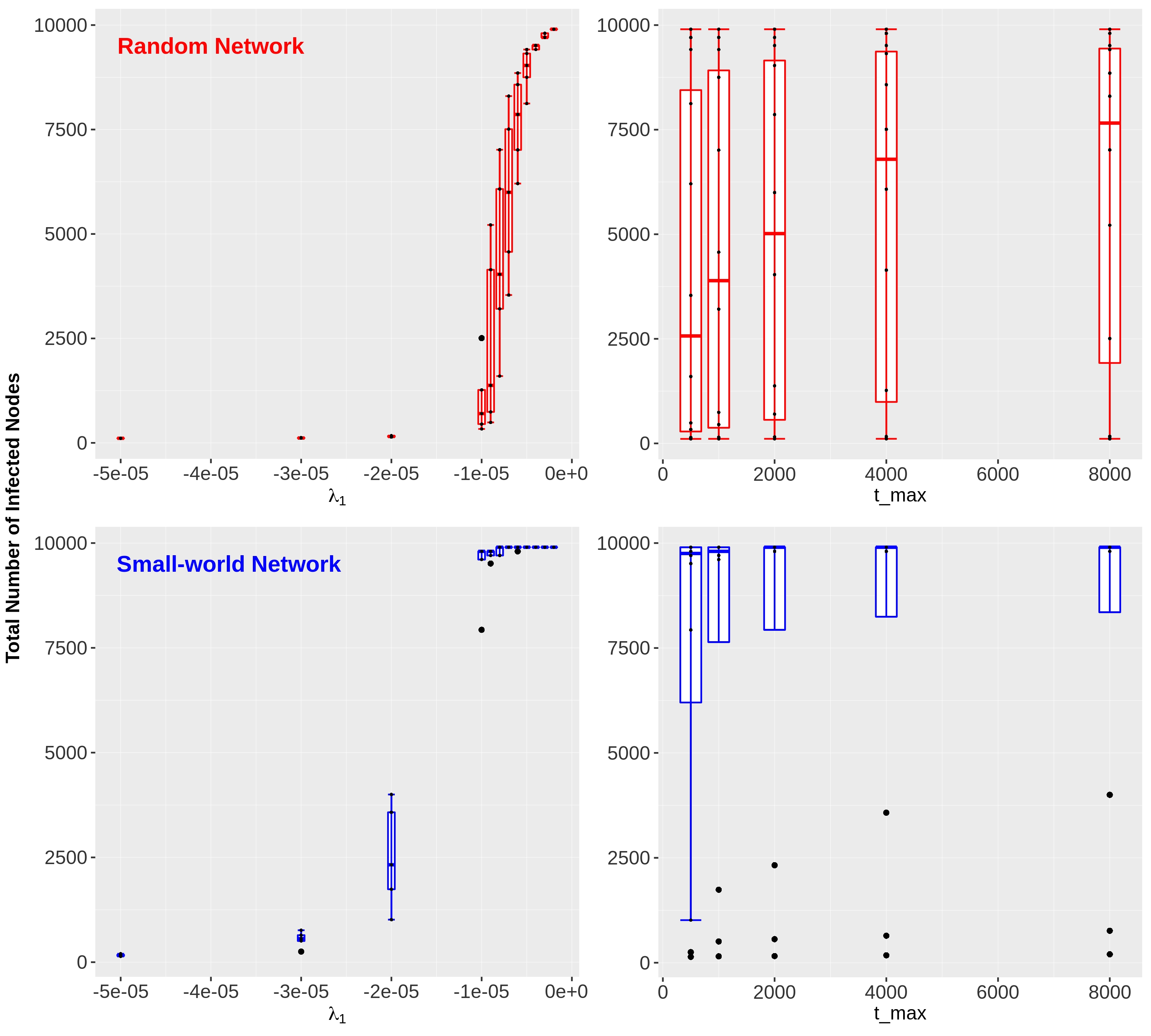}
\caption{Box plots of the number of infected nodes in 100 simulated cascades for Random network and Small-world network}
\vspace{-4mm}
\end{figure}

\begin{figure*}
\includegraphics[width=\textwidth]{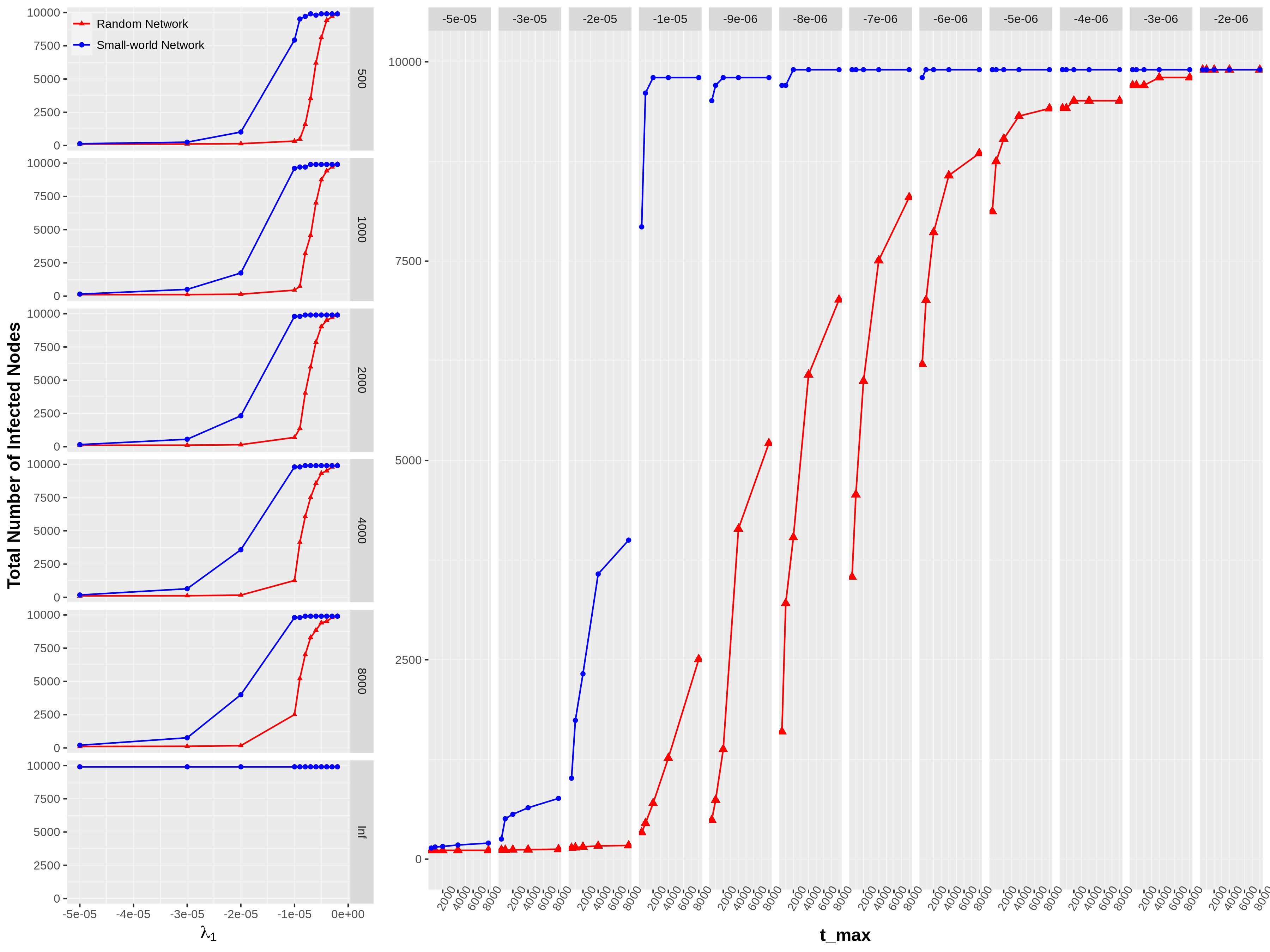}
\caption{Number of infected nodes in 100 simulated cascades for the random networks and small-world networks}
\vspace{-4mm}
\end{figure*}

Several conclusions about the influence of $\lambda_1$ and $t_{max}$ and the difference between the random network structure and the small-world network structure are drawn from these statistical results:
\begin{enumerate}
    \item If the cutoff diffusion time is set as $\infty$ where all nodes can get infected (i.e., the bottom row of the left plot in Figure 4, the value of $\lambda_1$ has no influence on the total number of infected nodes;
    \item For other cutoff diffusion times, it is clear that the geographic distance has a great influence on the total number of infected nodes. The number of infected nodes monotonically increases to the peak when $\lambda_1$ increases;
    \item When the \textbf{absolute value} of $\lambda_1$ is low, the influence of the distance on the diffusion is small, which results in an approximate constant hazard rate for all location-pairs;
    \item When the \textbf{absolute value} of $\lambda_1$ is high, more nodes will survive from contagions due to the geographic constraint as first several columns of the right plot in Figure 4;
    \item The maximum number of infected nodes that the small-world network can reach is more than that in the random network because the small-world network is highly clustered that makes the contagion spreads more easily;
    \item The simulation process of the random network is more sensitive to the change of parameters and the number of infected nodes will increase gradually when $\lambda_1$ and $t_{max}$ increase. This number in the small-world network will increase dramatically to the peak within a short range of $\lambda_1$ and is less sensitive to the change of the cutoff diffusion time except when $\lambda_1 = -2*10^{-5}$.
\end{enumerate}

When $\lambda_1$ is very small, most nodes fail to spread contagions for different cutoff diffusion time except for $\infty$. When $\lambda_1$ increases, most nodes in the small-world network can get infected within a short time due to an average small distance. So the influence of the cutoff diffusion time is only obvious for $\lambda_1$ around $-2*10^{-5}$. The distribution of the number of infected nodes of the random network is nearly normally distributed while that of the small-world network is more skewed with a large proportion of outliers identified.

\section{Visualization of Diffusion Cascades}
To better understand the geographic distribution of the infected nodes in the diffusion process, Figure 6 below shows simulated cascades generated by taking \textbf{Chicago} as the starting infected node under different sets of $\lambda_1$ and cutoff diffusion time $t_{max}$ for the random network and the small-world network. \par

\begin{figure*}[!ht]
\begin{multicols}{2}
\includegraphics[width=\linewidth]{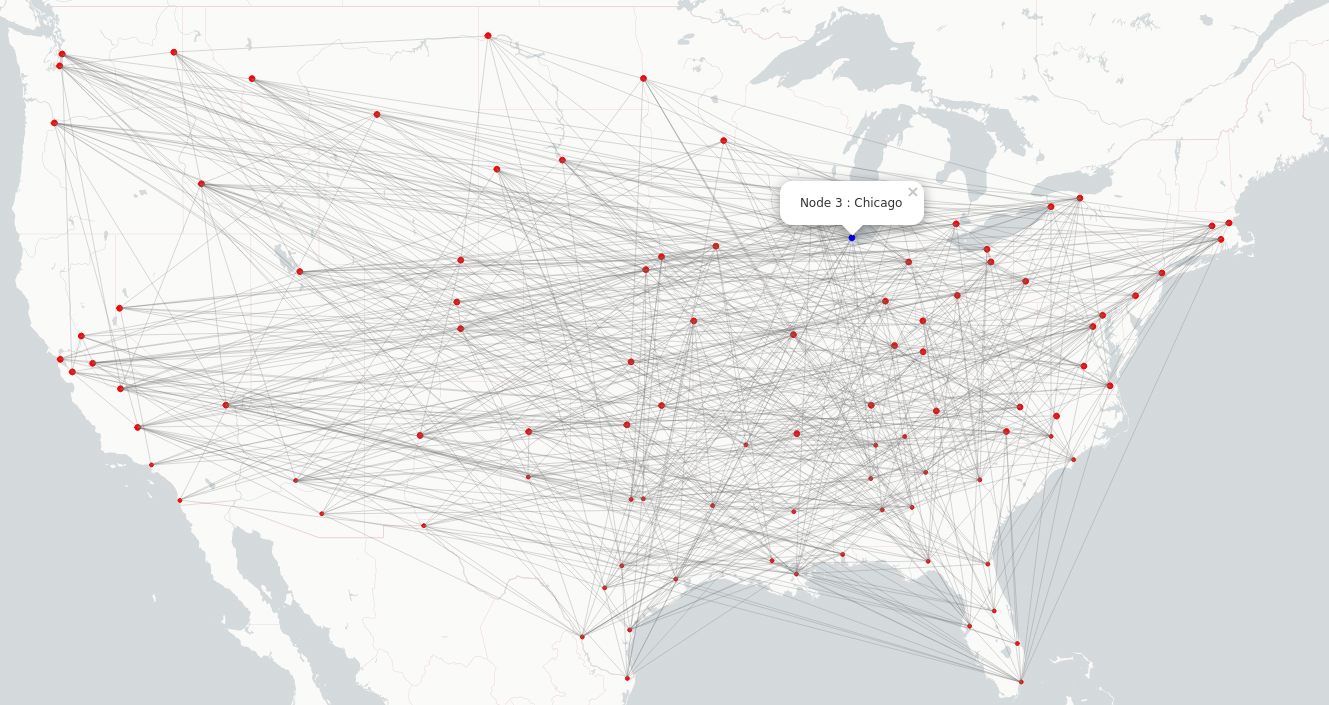}\hfill
\centering{$\lambda_1= -3*10^{-5}$, $t_{max} =1000$}
\includegraphics[width=\linewidth]{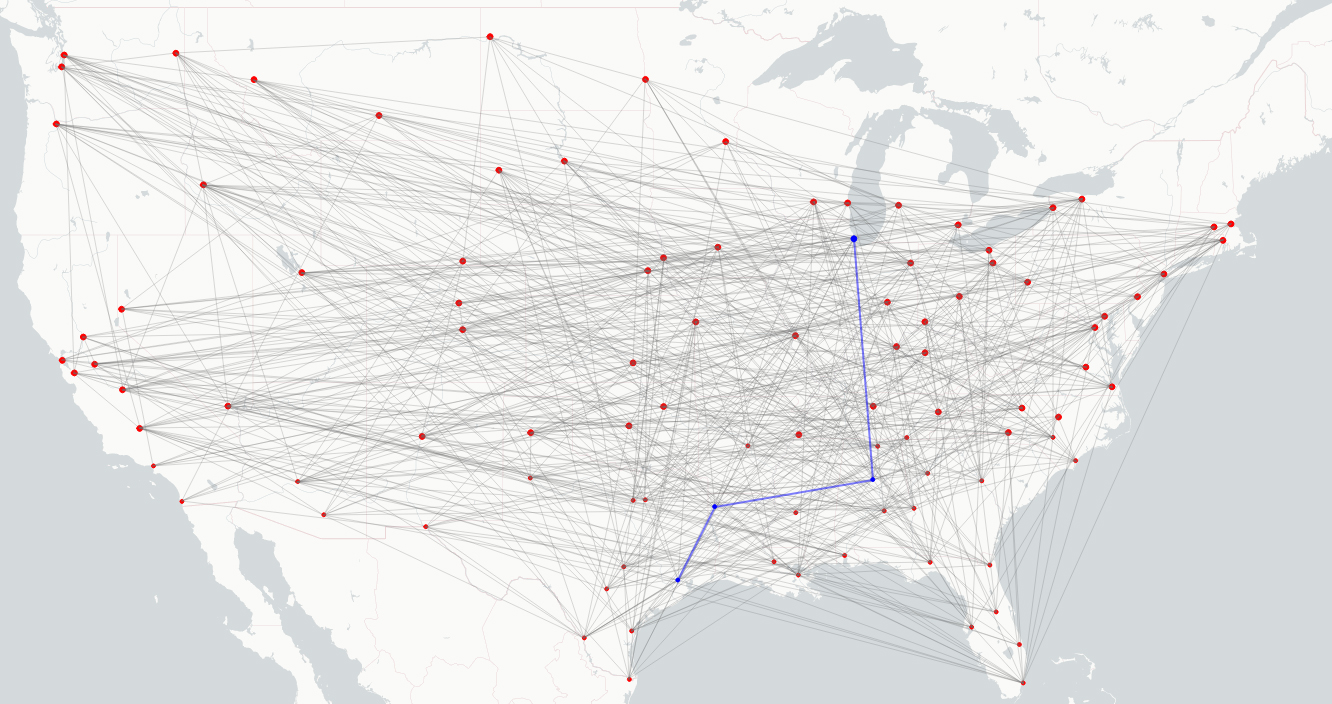}\hfill
\centering{$\lambda_1= -10^{-5}$, $t_{max} =1000$}
\end{multicols}
\begin{multicols}{2}
\includegraphics[width=\linewidth]{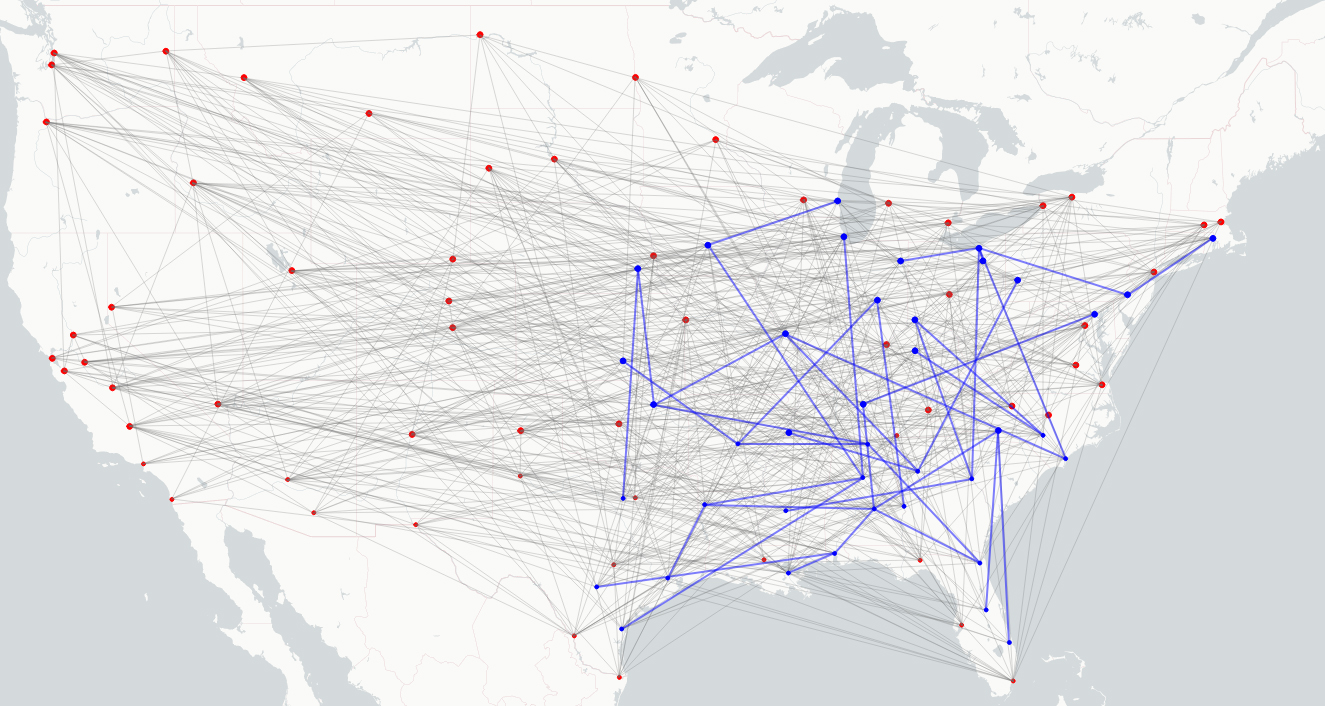}\hfill
\centering{$\lambda_1= -10^{-5}$, $t_{max} =4000$}
\includegraphics[width=\linewidth]{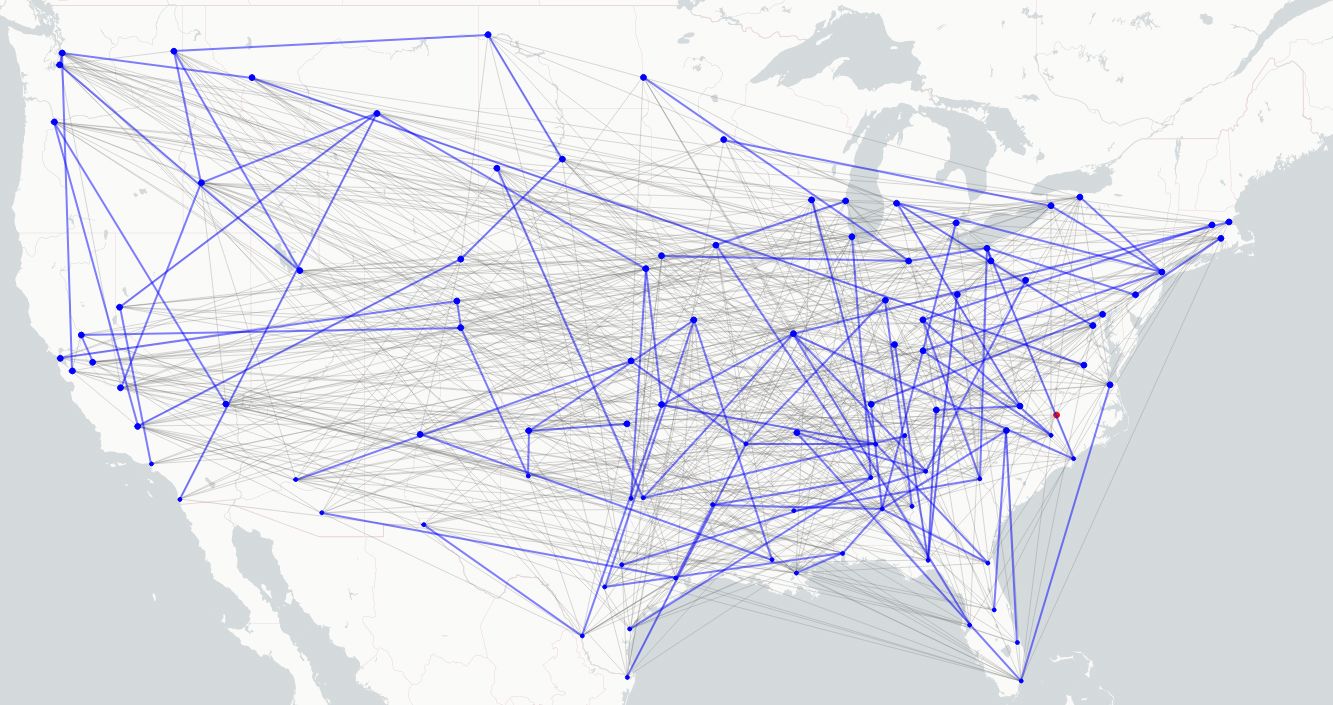}\hfill
\centering{$\lambda_1= -10^{-5}$, $t_{max} = \infty$}
\end{multicols}
\begin{multicols}{2}
\includegraphics[width=\linewidth]{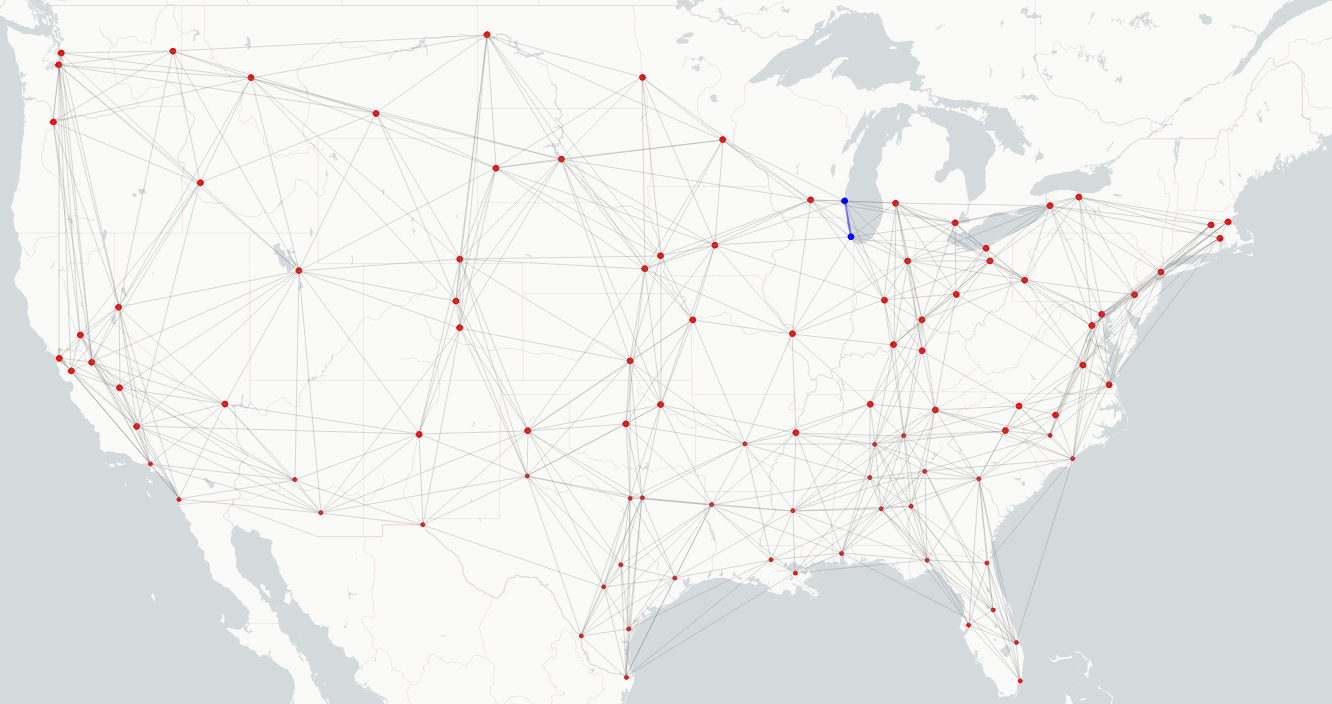}\hfill
\centering{$\lambda_1= -5*10^{-5}$, $t_{max} =500$}
\includegraphics[width=\linewidth]{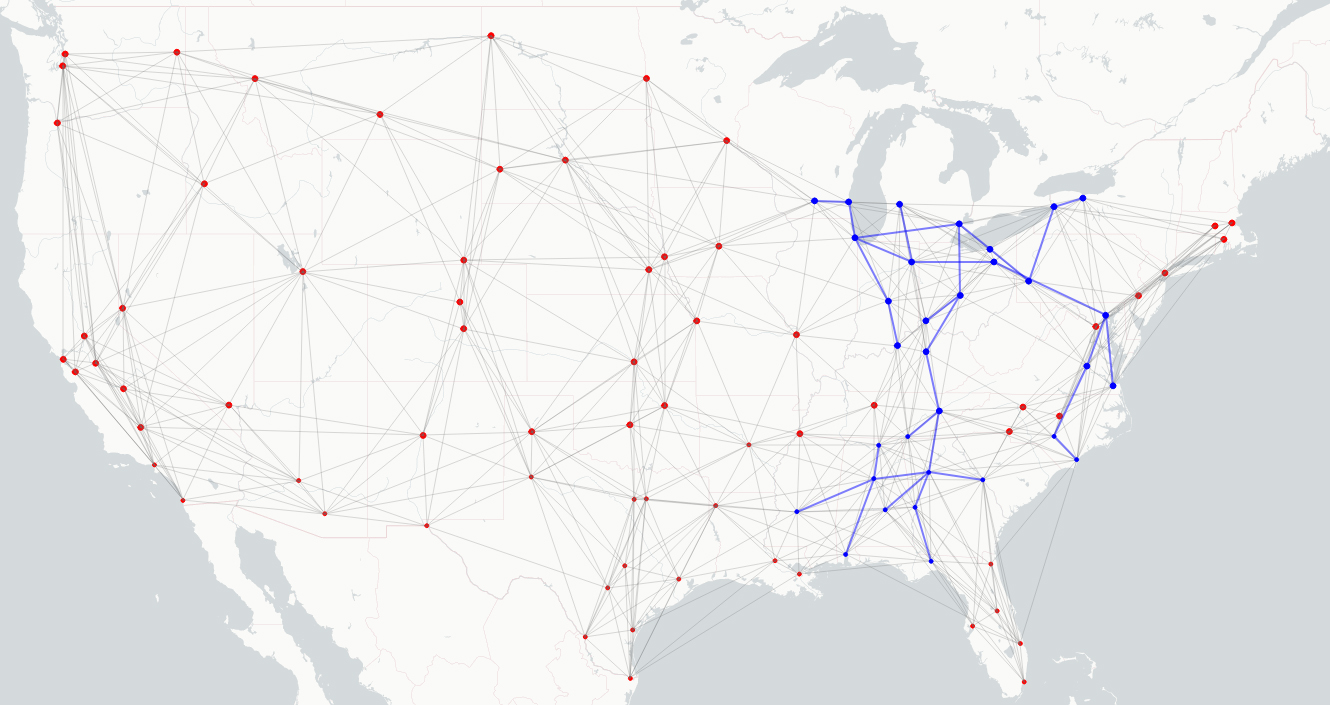}\hfill
\centering{$\lambda_1= -2*10^{-5}$, $t_{max} =500$}
\end{multicols}
\begin{multicols}{2}
\includegraphics[width=\linewidth]{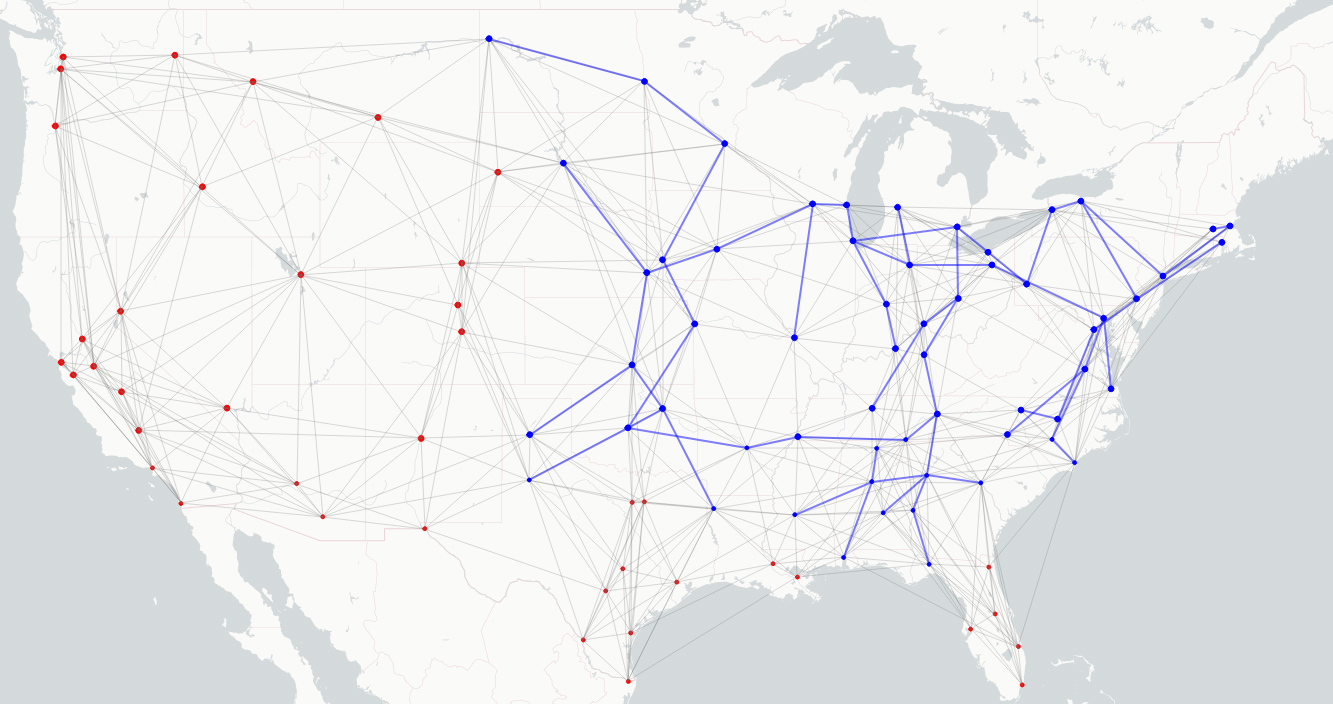}\hfill
\centering{$\lambda_1= -2*10^{-5}$, $t_{max} =8000$}
\includegraphics[width=\linewidth]{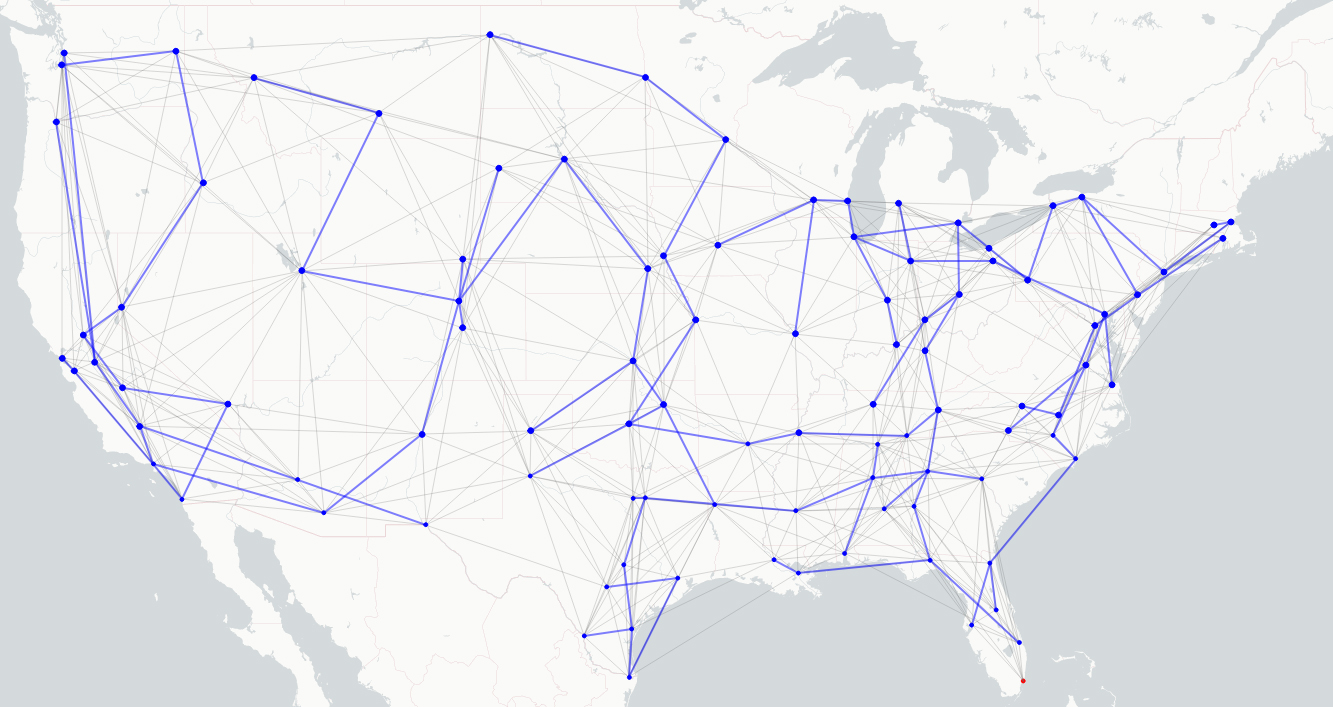}\hfill
\centering{$\lambda_1= -2*10^{-5}$, $t_{max} =\infty$}
\end{multicols}
\vspace{-7mm}
\caption{Simulated cascades in the random and small-world networks under different sets of $\lambda_1$ and $t_{max}$}
\end{figure*}

When $\lambda_1$ and cutoff diffusion time $t_{max}$ is very small, it is hard for contagions to spread over the random network. However, contagions can still spread from Chicago to Milwaukee in the small-world network when $\lambda_1= -5*10^{-5}$ and $t_{max}=500$ due to nodes in the small-world network are highly connected with a small average geographic distance. It is also intuitive to observe that the contagion spreads like a \textbf{chain via the connected cliques} in the small-world network but more \textbf{divergent} over space in the random network. When the cutoff diffusion time is $Inf$, still the node 40 (i.e., Raleigh) in the random network and the node 41 (i.e., Miami) in the small-world network haven't been infected. This is because some nodes only have outward connections. Contagions cannot spread to these nodes unless they are the starting infected node.

\section{Discussion}
First novel aspect of the STAND algorithm is that the geographic distance is implemented into the hazard rate function, \textbf{integrated over the time} to quantify its influence on the diffusion probability in Equation (9). By varying its coefficient $\lambda_1$, we can model different real-world diffusion phenomena, where the geographic aspects have different effects.  \par

The second novel aspect is that it can be applied to any kind of network structures that are observed in the real-world diffusion phenomena. The network structure is regarded as the input, and the hazard rate and diffusion probability are calculated separately for each single edge dependent on the distance and the infection time difference between two nodes that this edge connects. This is more flexible and accurate than previous research that assume a uniform global diffusion rate over the network. \par

The third novel aspect of the STAND algorithm is that the node properties and spatial dependencies (e.g., geographic distance) are considered to be vectors of explanatory variables $X$ and $Y$ in the definition of the hazard rate in Equation (6). Our STAND algorithm allows different node properties or spatial dependencies for different node-pairs within the same network since the diffusion probability is calculated separately for each edge.

\section{Conclusions and Future Work}
This research presents a network diffusion algorithm, called STAND, based on the framework of probabilistic survival modeling, to simulate the diffusion cascades and infer the underlying spatiotemporal network over which various contagions (e.g., political policies, social opinions, news, diseases, etc.) spread over both space and time. It can estimate the diffusion speed and the \textbf{individual-level} infection likelihood given any network structures as input. It can help discover the diffusion pathways and also predict new occurrences of infections over space and time as well as as well as trace back to detect where and when the diffusion began. \par

Real-world network diffusion phenomena are usually more complex than the simulated scenarios we have described here. The complexities can be summarized as follows: 1) the real-world networks over which contagions spread are a mixture of different levels of randomness and clustering rather than the simple random network or small-world network structure; and 2) the network nodes are heterogeneous and their properties have high influence on the diffusion process. We will address these complexities as the next step in our research. Future work will collect large-scale real-world spatiotemporal diffusion dataset and take various geographic characteristics into consideration, such as the local population or geographic scale.\par

Also, a more complex interactive visualization tool and an R package that integrates the STAND algorithm should be developed in the future to model the synthetic spatial network and simulated diffusion cascades as an aid to understanding high-dimensional and complex results. 

\section*{Data Availability}
We have published our codes and data at the \href{https://scholarsphere.psu.edu/collections/08k71nh108}{Spatiotemporal Network Diffusion Codes and Data} repository on the ScholarSphere. This repository collection includes: 1) 100 geographic locations we used for modeling the network nodes; 2) R codes for simulating both the random network and small-world network as well as the diffusion cascades spreading over these two network structures, using the proposed STAND algorithm; and 3) CSV files of simulated diffusion cascades under different sets of parameters $\lambda_1$ and $t_{max}$ \par

\begin{acks}
We would also like to thank to Prof. Guido Cervone who provided many suggestions and comments about this research.
\end{acks}

\bibliographystyle{ACM-Reference-Format}
\bibliography{acmart}


\begin{thebibliography}{14}


\ifx \showCODEN    \undefined \def \showCODEN     #1{\unskip}     \fi
\ifx \showDOI      \undefined \def \showDOI       #1{#1}\fi
\ifx \showISBNx    \undefined \def \showISBNx     #1{\unskip}     \fi
\ifx \showISBNxiii \undefined \def \showISBNxiii  #1{\unskip}     \fi
\ifx \showISSN     \undefined \def \showISSN      #1{\unskip}     \fi
\ifx \showLCCN     \undefined \def \showLCCN      #1{\unskip}     \fi
\ifx \shownote     \undefined \def \shownote      #1{#1}          \fi
\ifx \showarticletitle \undefined \def \showarticletitle #1{#1}   \fi
\ifx \showURL      \undefined \def \showURL       {\relax}        \fi
\providecommand\bibfield[2]{#2}
\providecommand\bibinfo[2]{#2}
\providecommand\natexlab[1]{#1}
\providecommand\showeprint[2][]{arXiv:#2}

\bibitem[\protect\citeauthoryear{Almquist}{Almquist}{2018}]%
        {almquist2018large}
\bibfield{author}{\bibinfo{person}{Zack~W Almquist}.}
  \bibinfo{year}{2018}\natexlab{}.
\newblock \showarticletitle{Large-scale spatial network models: An application
  to modeling information diffusion through the homeless population of {San
  Francisco}}.
\newblock \bibinfo{journal}{\emph{Environment and Planning B: Urban Analytics
  and City Science}} (\bibinfo{year}{2018}).
\newblock


\bibitem[\protect\citeauthoryear{Barrett, Bisset, Eubank, Feng, and
  Marathe}{Barrett et~al\mbox{.}}{2008}]%
        {barrett2008episimdemics}
\bibfield{author}{\bibinfo{person}{Christopher~L Barrett},
  \bibinfo{person}{Keith~R Bisset}, \bibinfo{person}{Stephen~G Eubank},
  \bibinfo{person}{Xizhou Feng}, {and} \bibinfo{person}{Madhav~V Marathe}.}
  \bibinfo{year}{2008}\natexlab{}.
\newblock \showarticletitle{EpiSimdemics: an efficient algorithm for simulating
  the spread of infectious disease over large realistic social networks}. In
  \bibinfo{booktitle}{\emph{SC'08: Proceedings of the 2008 ACM/IEEE Conference
  on Supercomputing}}. IEEE, \bibinfo{pages}{1--12}.
\newblock


\bibitem[\protect\citeauthoryear{Erdos and R{\'e}nyi}{Erdos and
  R{\'e}nyi}{1960}]%
        {erdos1960evolution}
\bibfield{author}{\bibinfo{person}{Paul Erdos} {and}
  \bibinfo{person}{Alfr{\'e}d R{\'e}nyi}.} \bibinfo{year}{1960}\natexlab{}.
\newblock \showarticletitle{On the evolution of random graphs}.
\newblock \bibinfo{journal}{\emph{Publication of the Mathematical Institute of
  the Hungarian Academy of Sciences}} \bibinfo{volume}{5}, \bibinfo{number}{1}
  (\bibinfo{year}{1960}), \bibinfo{pages}{17--60}.
\newblock


\bibitem[\protect\citeauthoryear{Kabir and Tanimoto}{Kabir and
  Tanimoto}{2019}]%
        {kabir2019analysis}
\bibfield{author}{\bibinfo{person}{KM~Ariful Kabir} {and} \bibinfo{person}{Jun
  Tanimoto}.} \bibinfo{year}{2019}\natexlab{}.
\newblock \showarticletitle{Analysis of epidemic outbreaks in two-layer
  networks with different structures for information spreading and disease
  diffusion}.
\newblock \bibinfo{journal}{\emph{Communications in Nonlinear Science and
  Numerical Simulation}} (\bibinfo{year}{2019}).
\newblock


\bibitem[\protect\citeauthoryear{Kiesling, G{\"u}nther, Stummer, and
  Wakolbinger}{Kiesling et~al\mbox{.}}{2012}]%
        {Kiesling2012}
\bibfield{author}{\bibinfo{person}{Elmar Kiesling}, \bibinfo{person}{Markus
  G{\"u}nther}, \bibinfo{person}{Christian Stummer}, {and}
  \bibinfo{person}{Lea~M. Wakolbinger}.} \bibinfo{year}{2012}\natexlab{}.
\newblock \showarticletitle{Agent-based simulation of innovation diffusion: a
  review}.
\newblock \bibinfo{journal}{\emph{Central European Journal of Operations
  Research}} (\bibinfo{year}{2012}).
\newblock


\bibitem[\protect\citeauthoryear{Leskovec and Horvitz}{Leskovec and
  Horvitz}{2014}]%
        {leskovec2014geospatial}
\bibfield{author}{\bibinfo{person}{Jure Leskovec} {and} \bibinfo{person}{Eric
  Horvitz}.} \bibinfo{year}{2014}\natexlab{}.
\newblock \showarticletitle{Geospatial structure of a planetary-scale social
  network}.
\newblock \bibinfo{journal}{\emph{IEEE Transactions on Computational Social
  Systems}} \bibinfo{volume}{1}, \bibinfo{number}{3} (\bibinfo{year}{2014}),
  \bibinfo{pages}{156--163}.
\newblock


\bibitem[\protect\citeauthoryear{Liben-Nowell and Kleinberg}{Liben-Nowell and
  Kleinberg}{2008}]%
        {liben2008tracing}
\bibfield{author}{\bibinfo{person}{David Liben-Nowell} {and}
  \bibinfo{person}{Jon Kleinberg}.} \bibinfo{year}{2008}\natexlab{}.
\newblock \showarticletitle{Tracing information flow on a global scale using
  {Internet} chain-letter data}.
\newblock \bibinfo{journal}{\emph{Proceedings of the National Academy of
  Sciences}} \bibinfo{volume}{105}, \bibinfo{number}{12}
  (\bibinfo{year}{2008}).
\newblock


\bibitem[\protect\citeauthoryear{Liben-Nowell, Novak, Kumar, Raghavan, and
  Tomkins}{Liben-Nowell et~al\mbox{.}}{2005}]%
        {liben2005geographic}
\bibfield{author}{\bibinfo{person}{David Liben-Nowell},
  \bibinfo{person}{Jasmine Novak}, \bibinfo{person}{Ravi Kumar},
  \bibinfo{person}{Prabhakar Raghavan}, {and} \bibinfo{person}{Andrew
  Tomkins}.} \bibinfo{year}{2005}\natexlab{}.
\newblock \showarticletitle{Geographic routing in social networks}.
\newblock \bibinfo{journal}{\emph{Proceedings of the National Academy of
  Sciences}} \bibinfo{volume}{102}, \bibinfo{number}{33}
  (\bibinfo{year}{2005}).
\newblock


\bibitem[\protect\citeauthoryear{Mahajan, Muller, and Wind}{Mahajan
  et~al\mbox{.}}{2000}]%
        {mahajan2000new}
\bibfield{author}{\bibinfo{person}{Vijay Mahajan}, \bibinfo{person}{Eitan
  Muller}, {and} \bibinfo{person}{Yoram Wind}.}
  \bibinfo{year}{2000}\natexlab{}.
\newblock \bibinfo{booktitle}{\emph{New-product diffusion models}}.
  Vol.~\bibinfo{volume}{11}.
\newblock \bibinfo{publisher}{Springer Science \& Business Media}.
\newblock


\bibitem[\protect\citeauthoryear{Rahmandad and Sterman}{Rahmandad and
  Sterman}{2008}]%
        {rahmandad2008heterogeneity}
\bibfield{author}{\bibinfo{person}{Hazhir Rahmandad} {and}
  \bibinfo{person}{John Sterman}.} \bibinfo{year}{2008}\natexlab{}.
\newblock \showarticletitle{Heterogeneity and network structure in the dynamics
  of diffusion: Comparing agent-based and differential equation models}.
\newblock \bibinfo{journal}{\emph{Management Science}} \bibinfo{volume}{54},
  \bibinfo{number}{5} (\bibinfo{year}{2008}), \bibinfo{pages}{998--1014}.
\newblock


\bibitem[\protect\citeauthoryear{Rodriguez, Balduzzi, and
  Sch{\"o}lkopf}{Rodriguez et~al\mbox{.}}{2011}]%
        {rodriguez2011uncovering}
\bibfield{author}{\bibinfo{person}{Manuel~Gomez Rodriguez},
  \bibinfo{person}{David Balduzzi}, {and} \bibinfo{person}{Bernhard
  Sch{\"o}lkopf}.} \bibinfo{year}{2011}\natexlab{}.
\newblock \showarticletitle{Uncovering the temporal dynamics of diffusion
  networks}.
\newblock \bibinfo{journal}{\emph{The 28th International Conference on Machine
  Learning}} (\bibinfo{year}{2011}).
\newblock


\bibitem[\protect\citeauthoryear{Taylor and Rowlingson}{Taylor and
  Rowlingson}{2017}]%
        {taylor2017spatsurv}
\bibfield{author}{\bibinfo{person}{Benjamin Taylor} {and}
  \bibinfo{person}{Barry Rowlingson}.} \bibinfo{year}{2017}\natexlab{}.
\newblock \showarticletitle{spatsurv: An {R} package for Bayesian inference
  with spatial survival models}.
\newblock \bibinfo{journal}{\emph{Journal of Statistical Software}}
  \bibinfo{volume}{77}, \bibinfo{number}{4} (\bibinfo{year}{2017}),
  \bibinfo{pages}{1--32}.
\newblock


\bibitem[\protect\citeauthoryear{Wang, Ye, and Tsou}{Wang
  et~al\mbox{.}}{2016}]%
        {wang2016spatial}
\bibfield{author}{\bibinfo{person}{Zheye Wang}, \bibinfo{person}{Xinyue Ye},
  {and} \bibinfo{person}{Ming-Hsiang Tsou}.} \bibinfo{year}{2016}\natexlab{}.
\newblock \showarticletitle{Spatial, temporal, and content analysis of Twitter
  for wildfire hazards}.
\newblock \bibinfo{journal}{\emph{Natural Hazards}} \bibinfo{volume}{83},
  \bibinfo{number}{1} (\bibinfo{year}{2016}), \bibinfo{pages}{523--540}.
\newblock


\bibitem[\protect\citeauthoryear{Watts and Strogatz}{Watts and
  Strogatz}{1998}]%
        {watts1998collective}
\bibfield{author}{\bibinfo{person}{Duncan~J Watts} {and}
  \bibinfo{person}{Steven~H Strogatz}.} \bibinfo{year}{1998}\natexlab{}.
\newblock \showarticletitle{Collective dynamics of small-world networks}.
\newblock \bibinfo{journal}{\emph{Nature}} \bibinfo{volume}{393},
  \bibinfo{number}{6684} (\bibinfo{year}{1998}).
\newblock


\end{thebibliography}

\end{document}